\documentclass[twocolumn,aps,prb,epsf,showpacs]{revtex4}

\usepackage{amsmath} 
\usepackage{amsfonts} 
\usepackage{epsfig} 
\usepackage{epsf} 
\usepackage{array}

\begin{document}

\title{Magnetic interaction at an interface between manganite and other transition-metal oxides} 
\author{Satoshi Okamoto}
\affiliation{Materials Science and Technology Division, Oak Ridge National Laboratory, Oak Ridge, Tennessee 37831-6071, USA}

\begin{abstract}
A general consideration is presented for the magnetic interaction at an interface 
between a perovskite manganite and other transition metal oxides. 
The latter is specified by the electron number $n$ in the $d_{3z^2-r^2}$ level as $(d_{3z^2-r^2})^n$. 
Based on the molecular orbitals formed at the interface and the generalized Hund's rule, 
the sign of the magnetic interaction is rather uniquely determined. 
The exception is when the $d_{3z^2-r^2}$ orbital is stabilized in the interfacial manganite layer 
neighboring to a $(d_{3z^2-r^2})^1$ or $(d_{3z^2-r^2})^2$ system. 
In this case, the magnetic interaction is sensitive to the occupancy of the Mn $d_{3z^2-r2}$ orbital. 
It is also shown that the magnetic interaction between the interfacial Mn layer and the bulk region can be changed. 
Manganite-based heterostructures thus show a rich magnetic behavior. 
We also present how to generalize the argument including $t_{2g}$ orbitals. 
\end{abstract}

\maketitle

\section{Introduction}
Transition metal (TM) oxides have been one of the main subjects of materials science for decades. 
Experimental and theoretical efforts are driven by their rich, complex, and potentially useful behaviors 
originating from strong correlations between electrons and/or electrons and lattices.\cite{Imada98}
The recent developments in the crystal growth techniques, in particular the (laser) molecular-beam epitaxy, 
have made us recognize the opportunity to further control their behaviors and 
to generate phenomena that are not realized in the bulk systems.%
\cite{Izumi01,Ohtomo02,Ohtomo04,Okamoto04,Stahn05,Chakhalian06,Hoffmann05,Chakhalian07,Bhattacharya08}

Here, we focus on the magnetic behavior at an interface between perovskite manganite and other TM oxides. 
Perovskite manganites, especially La$_{1-x}$Sr$_x$MnO$_3$ (LSMO), 
are particularly important because of their ferromagnetic (F) metallic behavior 
with relatively high Curie temperature $T_C$ and large polarization. 
Controlling the magnetic interaction at interfaces involving manganites would cause a technological breakthrough for 
electronic devices using, for example, a tunneling magnetoresistance (TMR) effect\cite{Bowen03,Ogimoto03} 
and an exchange bias (EB) effect.\cite{Yu09}
This requires the microscopic information on the orbital states, not only on the spin states  
as demonstrated for cuprate/manganite interfaces in Refs.~\onlinecite{Chakhalian07} and \onlinecite{Veenendaal08}. 
However, it remains controversial whether the magnetic moment is induced in the cuprate region\cite{Chakhalian06,Veenendaal08}
or the dead layers appear in the manganite region.\cite{Hoffmann05,Luo08} 

The difficulty dealing with interfaces involving strongly-correlated electron systems comes from 
the small volume fraction which makes the experimental analysis challenging, and 
strong-correlation effects which hinder some of theoretical treatments. 
Therefore, if a Goodenough-Kanamori-type \cite{Goodenough63,Kanamori59}
transparent description of the interfacial magnetic interaction becomes available, 
both experiment and theory would greatly benefit.

In this paper, we present a general consideration for the magnetic interaction at an interface involving manganites. 
We first focus on the interfacial interaction derived by $d_{3z^2-r^2}$ orbitals 
which have the largest hybridization along the $z$ layer-stacking direction. 
We see that the sign of the magnetic interaction via the $d_{3z^2-r^2}$ orbitals is naturally fixed 
based on the molecular orbitals formed at the interface and the generalized Hund's rule. 
The argument uses localized orbitals, and therefore shows only the qualitative trend. 
The molecular orbitals effectively lift the degeneracy between $d_{3z^2-r^2}$ and $d_{x^2-y^2}$ orbitals 
by the order of the hopping intensity. 
In the second part, we perform the model Hartree-Fock calculation and show that the broken degeneracy 
can lead to the additional change in the magnetic interaction between the interfacial Mn layer and its neighboring Mn layer. 
We also discuss how to generalize the molecular-orbital based argument for more complicated situations including $t_{2g}$ orbitals.

\section{Molecular-orbital picture}

In this section, we consider the magnetic interaction between manganite and other TM oxides focusing on 
the molecular orbitals formed by $d_{3z^2-r^2}$ orbitals which have the largest overlap at the interface. 
The TM region is specified by the number of electrons occupying a $d_{3z^2-r^2}$ orbital. 
Here, $t_{2g}$ electrons are assumed to be electronically inactive and considered as localized spins 
when finite number of electrons occupy $t_{2g}$ orbitals. 
Generalization including these electrons will be discussed later. 

\begin{figure}[tbp] 
\includegraphics[width=0.6\columnwidth,clip]{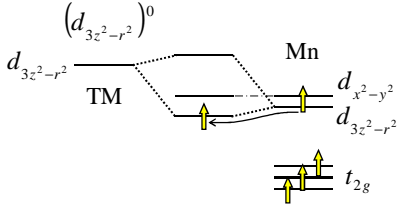} 
\caption{(Color online) Molecular orbital (middle) formed by $d_{3z^2-r^2}$ orbitals on Mn (right) and 
TM with the $(d_{3z^2-r^2})^0$ configuration (left). }
\label{fig:d0} 
\end{figure}

{\em $(d_{3z^2-r^2})^0$ system}. 
Let us start from the simplest case, 
an interface between Mn and a $(d_{3z^2-r^2})^0$ system (Fig.~\ref{fig:d0}). 
In this case, the bonding (B) orbital is occupied by an electron whose spin is parallel to the localized $t_{2g}$ spin in Mn, 
while the antibonding (AB) orbital is unoccupied. 
When there are other unpaired electrons in the $(d_{3z^2-r^2})^0$ system at $d_{x^2-y^2}$ and/or $t_{2g}$ orbitals, 
their spins align parallel to that of the electron in the B orbital due to the Hund coupling. 
Thus, the F coupling is generated between Mn and $(d_{3z^2-r^2})^0$ systems. 
This is equivalent to the double-exchange (DE) interaction originally proposed by Zener.\cite{Zener51} 
When $d_{x^2-y^2}$ is much lower in energy than $d_{3z^2-r^2}$ and in the interfacial Mn layer (termed $d_{x^2-y^2}$ order), 
Mn and $(d_{3z^2-r^2})^0$ systems are virtually decoupled. 
Thus, the magnetic coupling is due to the superexchange (SE) interaction between $t_{2g}$ electrons. 
This interaction is either F or antiferromagnetic (AF) depending on the orbital state and the occupancy of the TM $t_{2g}$ level.

\begin{figure}[tbp] 
\includegraphics[width=0.8\columnwidth,clip]{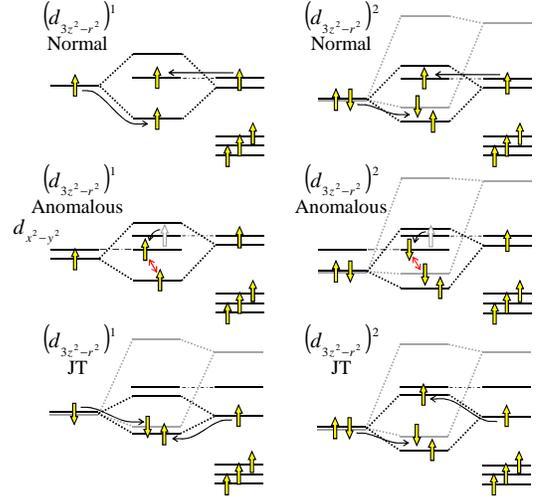} 
\caption{(Color online) Molecular orbitals formed by $d_{3z^2-r^2}$ orbitals on Mn and 
the $(d_{3z^2-r^2})^1$ system (left column) and the $(d_{3z^2-r^2})^2$ system (right column). 
In the normal (anomalous) configurations, $d_{3z^2-r^2}$ and $d_{x^2-y^2}$ orbitals are nearly degenerate in the interfacial Mn, 
and the unoccupied $d_{x^2-y^2}$ orbital in the neighboring TM is higher in energy (lower in energy than the occupied Mn $d_{x^2-y^2}$). 
In the JT case, the $d_{3z^2-r^2}$ level is significantly lower than the $d_{x^2-y^2}$ level. 
Black (light) lines indicate the level of majority (minority) spins. 
The up level and down level are exchange split resulting in the level scheme as indicated.
The minority levels are neglected in the upper left two because these are irrelevant.} 
\label{fig:d12} 
\end{figure}

{\em $(d_{3z^2-r^2})^{1,2}$ normal}. 
This simple consideration can be easily generalized to $(d_{3z^2-r^2})^1$ and $(d_{3z^2-r^2})^2$ systems. 
First we consider that the $d_{3z^2-r^2}$ and $d_{x^2-y^2}$ are nearly degenerate in the interfacial Mn layer and 
the unoccupied $d_{x^2-y^2}$ level in the TM region is much higher than the $d_{3z^2-r^2}$ level 
(Fig.~\ref{fig:d12}, top figures). 
We call this configuration ``normal'' (N) configuration. 
In the lowest energy configuration, B orbitals and the Mn $d_{x^2-y^2}$ orbital are occupied by electrons. 
For the $(d_{3z^2-r^2})^1$ system, the F interaction is favorable as in the $(d_{3z^2-r^2})^0$ system. 
On the other hand, for the $(d_{3z^2-r^2})^2$ system, the down electron orbital is hybridized with the minority band in the Mn region. 
Thus, the ``down'' B orbital is higher in energy and has larger weight on the TM than the ``up'' B orbital. 
Because of the Hund coupling with the ``down'' electron in the B orbital, 
other unpaired electrons, if they exist in $d_{x^2-y^2}$ and/or $t_{2g}$ orbitals, 
tend to be antiparallel to the Mn spin. 

Since $d_{x^2-y^2}$ orbitals are predominantly occupied in the interfacial Mn layer 
due to the B/AB splitting of $d_{3z^2-r^2}$-based molecular orbitals, 
further stabilization of $d_{x^2-y^2}$ orbitals in the Mn region, i.e., $d_{x^2-y^2}$ order, 
would not affect the interfacial magnetic coupling discussed here. 
But, this could reverse the magnetic coupling between the interfacial Mn layer and the second Mn layer as discussed in the next section.

{\em $(d_{3z^2-r^2})^{1,2}$ anomalous}. 
When the $d_{x^2-y^2}$ level in the TM region becomes lower than the Mn $d_{x^2-y^2}$ level, the charge transfer occurs. 
We shall call this configuration ``anomalous'' (AN) configuration (Fig.~\ref{fig:d12}, middle figures). 
The electron transferred to the TM $d_{x^2-y^2}$ orbital has the same spin 
as the higher energy B orbital due to the Hund coupling (indicated by arrows). 
Therefore, the sign of the magnetic coupling between the Mn and $(d_{3z^2-r^2})^{1,2}$ systems is unchanged. 

Note that this argument is applicable when the hopping probability between $d_{x^2-y^2}$ orbitals in 
the Mn and the TM regions is negligibly small. 
The finite hopping probability would make the charge transfer continuous. 
Furthermore, when the hopping probability becomes large, the DE interaction is generated. 
Although the DE interaction through $d_{x^2-y^2}$ bonds may not be realistic, 
it cooperatively stabilizes the F spin alignment for the $(d_{3z^2-r^2})^1$ case, 
while it competes with the AF tendency for the $(d_{3z^2-r^2})^2$ case. 

Magnetic interactions discussed so far are insensitive to the electron density in the interfacial Mn 
because the interactions are mainly derived from the virtual electron excitation from the occupied $d_{3z^2-r^2}$ orbital in the TM region 
to the unoccupied counterpart in the Mn region. 

Next, we consider that the $d_{3z^2-r^2}$ level is much lower than the $d_{x^2-y^2}$ in the interfacial Mn layer 
due to either the local Jahn-Teller distortion or compressive strain originating from the substrate
(Fig.~\ref{fig:d12}, lower figures denoted by ``JT''). 
The magnetic coupling in this case is sensitive to the electron density of the interfacial Mn. 

{\em $(d_{3z^2-r^2})^1$ JT}. 
When the Mn $d_{3z^2-r^2}$ density is close to 1, the SE interaction between the occupied $d_{3z^2-r^2}$ orbitals becomes 
AF. 
On the other hand, when the density is much less than 1, the F interaction between $(d_{3z^2-r^2})^0$ configuration on Mn and 
$(d_{3z^2-r^2})^1$ becomes dominant. 

{\em $(d_{3z^2-r^2})^2$ JT}. 
When the Mn $d_{3z^2-r^2}$ occupancy is close to 1, 
``up'' electrons are localized on each sites because both B and AB molecular orbitals are occupied while down electrons can be excited or leaked 
from the $(d_{3z^2-r^2})^2$ system to the Mn minority level, i.e., down electron density is virtually reduced in the TM region. 
As a result, unpaired spins, if they exist in $d_{x^2-y^2}$ and/or $t_{2g}$ orbitals, 
become parallel to the up spin, i.e., F coupling. 
When the Mn $d_{3z^2-r^2}$ density becomes much less than 1, the up AB orbital becomes less occupied while keeping the occupancy of
B orbitals relatively unchanged. 
Eventually, the down density in the TM region becomes larger than the up density, 
and the magnetic coupling between the Mn and $(d_{3z^2-r^2})^2$ regions becomes AF.

\section{Model Hartree-Fock analysis}

In the previous section, we discussed the interfacial magnetic coupling controlled by the molecular orbitals. 
The separation between B and AB molecular levels can become as large as the order of 
$t$, the hybridization between $d_{3z^2-r^2}$ orbitals along the $z$ direction. 
Since the interfacial Mn $d_{3z^2-r^2}$ band is represented by the B (AB) $d_{3z^2-r^2}$ orbital 
for a $(d_{3z^2-r^2})^{0(1,2)}$/manganite interface, 
the $e_g$ degeneracy is effectively lifted in the interface layer. 
This degeneracy lifting is expected to affect the magnetic interaction in the Mn region. 
In this section, we discuss this effect using the microscopic model calculation. 

We consider a two-band DE model given by 
\begin{eqnarray}
H \!\!&=&\!\! \sum_{i} \Delta n_{i \alpha} 
-\sum_{\langle ij \rangle a b} \bigl\{ t_{ij}^{ab} U_{ij} d_{i a}^\dag d_{j b} + h.c. \bigr\} \nonumber \\
&& + \sum_i \tilde U n_{i \alpha} n_{i \beta} + J \sum_{\langle ij \rangle} \vec S_{t i} \cdot \vec S_{t j}. 
\label{eq:Hamiltonian}
\end{eqnarray}
Here, an electron annihilation operator at site $i$ and orbital $a [= \alpha (d_{3z^2-r^2}),\beta (d_{x^2-y^2})]$ is given by $d_{i a}$, 
$n_{i a} = d_{ia}^\dag d_{ia}$, and the level difference between $\alpha$ and $\beta$ is given by $\Delta$.  
We consider the large Hund coupling limit, in which 
the spin direction of a conduction electron is always parallel to that of a localized $t_{2g}$ spin on the same site, 
and omit the spin index. 
Instead, the relative orientation of $t_{2g}$ spins is reflected in the hopping matrix; 
$U_{ij}$ is the unitary transformation representing the rotation of the spin direction between sites $i$ and $j$. 
For simplicity, we only consider nearest-neighboring (NN) hoppings between Mn $e_g$ orbitals via oxygen $2p$ in the middle. 
Using the Slater-Koster scheme,\cite{Slater54}
the orbital dependence of $t_{ij}^{ab}$ is written as 
$t_{i, i+z}^{\alpha \alpha} = 4 t_{i, i+x(y)}^{\alpha \alpha} = t$, 
$t_{i, i+x(y)}^{\beta \beta}= \frac{3t}{4}$, 
$t_{i, i+x(y)}^{\alpha \beta}= t_{i, i+x(y)}^{\beta \alpha} = (-)\frac{\sqrt{3}t}{4}$, and 
$t_{i, i+z}^{\alpha \beta, \beta \alpha}=t_{i, i+z}^{\beta \beta}= 0$. 
The third term represents interorbital Coulomb interaction. 
Due to the $e_g$ symmetry, $\tilde U$ is related to the intraorbital Coulomb interaction $U$ and 
the interorbital exchange integral $J_H$ as $\tilde U = U- 3J_H$. 
The last term represents the AF SE interaction between NN $t_{2g}$ spins $|\vec S_t|=\frac{3}{2}$. 

%
From the optical measurements, the on-site interactions are estimated as $U \sim 3$~eV and $J_H \sim 0.5$~eV.\cite{Kovaleva04}
The density-functional theory calculation provides $t \sim 0.5$~eV.\cite{Ederer07}
Using the mean-field analysis for the N{\'e}el temperature $\sim 120K$ of CaMnO$_3$, 
one estimates $J \sim 1$~meV.\cite{Wollan55} 
A similar value is obtained from the magnon excitations in the A-AF phases of 
50~\% doped Pr$_{1-x}$Sr$_x$MnO$_3$ and Nd$_{1-x}$Sr$_x$MnO$_3$ 
supposing that the AF interaction is due to the same $J$.\cite{Kawano03} 
Thus, in what follows, we take $\tilde U=3t$. 
Considering some ambiguity, the realistic value for $JS_t^2/t$ is expected to be 
$\sim 0.01$--0.05.

We analyze the model Hamiltonian, Eq.~(\ref{eq:Hamiltonian}), using the Hartree-Fock approximation at $T=0$ 
focusing on the doped region (carrier density $N$ far away from 1). 
In light of the experimental reports, 
we compare the energy of the following eight magnetic orderings: 
F ordering, 
planar AF ordering in which spins align ferromagnetically in the $xy$ ($xz$ or $yz$) plane [A (A')], 
chain-type AF ordering in which spins align ferromagnetically along the $z$ ($x$ or $y$) direction [C (C')], 
zigzag AF in which spins form ferromagnetic zigzag chains in the $xy$ ($xz$ or $yz$) plane [CE (CE')], 
and NaCl-type AF (G). 
At $N \rightarrow 1$, in addition to the spin symmetry breaking, orbital symmetry can be broken 
due to the SE mechanism in the present model.\cite{Kugel82} 
Since we are focusing on the metallic regime $N<1$, we do not consider such a symmetry breaking.

\begin{figure}[tbp] 
\includegraphics[width=0.8\columnwidth,clip]{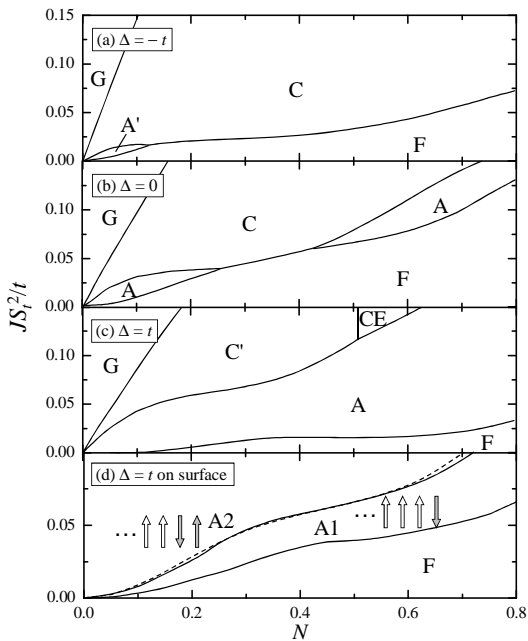} 
\caption{[(a)-(c)] Mean-field phase diagrams of doped manganites as a function of electron density $N$ 
and the AF interaction $J$ 
for three choices of the level difference between $e_g$ orbitals $\Delta$. 
At $\Delta< (>) 0$, $d_{3z^2-r^2}$ is lower (higher) in energy than $d_{x^2-y^2}$ . 
For notations of the magnetic phases, see the main text. 
(d) Phase diagram for the 20-layer slab with $\Delta=t$ in the surface layers. 
$N$ in this case corresponds to the mean electron density. 
For A1 and A2 phases, schematic spin alignments are also shown. 
A dashed line is the phase boundary between F and A-AF phases in the bulk calculation. }
\label{fig:pd} 
\end{figure}

The numerical results for the bulk phase diagram are presented in Figs.~\ref{fig:pd}~(a)-\ref{fig:pd}~(c). 
Here, all phase boundaries are of first order, and 
those at small $N$ can be replaced by canted AF phases or the phase separation between the undoped G-AF phase and doped F or AF phases. 
The overall feature is consistent with the previous theoretical reports.\cite{Maezono98,Brink99a,Fang00} 
At $\Delta=0$, A-AF and A'-AF (C- and C'-, CE- and CE'-) are degenerate but the degeneracy is lifted by the finite $\Delta$. 
We found the CE phase at $\tilde U=\Delta = 0$ at $JS_t^2 \agt 0.112 t$ and $N \agt 0.5$ (not shown) 
as in the previous reports.\cite{Solovyev99,Brink99b,Brey05} 
At $\tilde U = 3t$, the CE phase becomes unstable against A- and C-AF phases 
and appears only at the positive $\Delta$ with $JS_t^2 \agt 0.1t$. 
$JS_t^2 \sim 0.05t$ reproduces the phase diagram of a high $T_C$ system 
such as LSMO and Pr$_{1-x}$Sr$_x$MnO$_3$ (Refs. \onlinecite{Jirak01} and \onlinecite{Chmaissem03} fairly well. 

The main effect of the level separation $\Delta$ is changing the stability of planar-type AF (A or A') with respect to the 
chain-type AF (C or C') and F states. 
In particular, the A-AF phase is stabilized by the positive $\Delta$ more strongly than the C-AF phase by the negative $\Delta$. 
This is because the energy gain by the DE mechanism is favorable for the A-AF than the C-AF. 
The result is semiquantitatively consistent with the previous report based on the density-functional theory.\cite{Fang00}
At $\Delta=t$, the boundary between F and A-AF phases is moved down to 
$JS_t^2 \sim 0.02t$ at $0.3 \alt N \alt 0.7$. 
This behavior suggests that, when the $d_{3z^2-r^2}$ AB level for the $(d_{3z^2-r^2})^{1,2}$/manganite interfaces 
is about $t$ higher than the Mn $d_{x^2-y^2}$ level, 
the magnetic coupling between the interfacial Mn and the second Mn layers is switched to AF while retaining the intraplane F coupling. 

We confirmed this behavior by computing the surface phase diagram considering F phase and two AF phases: 
A1(2) where the surface layer (and the second layer) is antiferromagnetically coupled to its neighbor. 
We introduce positive $\Delta$ only on the surface layers in the 20-layer slab. 
As shown in Fig.~\ref{fig:pd} (d), a large part of F phase is replaced by A1 phase compared with the bulk phase diagram (b). 
(Precise phase boundary requires detailed information of the surface or interface.)
Although the parameter regime is small, 
it is also possible that surface three layers are AF coupled while the other couplings remain F, A2 phase, 
before the whole system enters A-AF when $J$ is increased or $N$ is decreased. 
When the $d_{3z^2-r^2}$ orbital is stabilized, 
the inter-Mn-layer coupling remains F but the intraplane F coupling is reduced. 
Therefore, in-plane canted AF structure may result for small $N$.

\section{Summary and discussion}

Summarizing, we presented a general consideration on the magnetic interaction between the doped manganite and other transition metal oxides 
when an interface is formed. 
Using the molecular orbital formed at the interface and the generalized Hund's rule, 
the sign of the magnetic interaction is determined (Sec. II). 
%
The bonding/antibonding splitting of the molecular orbitals leads to the degeneracy lifting of $e_g$ orbitals on the interface Mn layer. 
Further, the bulk strain lifts the $e_g$ degeneracy. 
These effects control the magnetic interaction in the interfacial Mn plane and between the interfacial Mn plane and its neighbor (Sec. III). 
Considering these effects, we summarized the magnetic couplings in $(d_{3z^2-r^2})^{n}$/manganite interfaces 
in Table~\ref{tab:coupling}. 
Although the present argument is rather qualitative, 
it is physically transparent and can be applied to a variety of systems. 
It is also straightforward to generalize the argument to include other orbitals. 
Therefore, the present argument will also help a more quantitative analysis 
with detailed information from either the experiment or the first principle theory.

\begin{table}[tbp]
\caption{Magnetic interaction at an interface between Mn and TM with the $(d_{3z^2-r^2})^n$ configuration. 
The interfacial Mn is indicated by Mn(1), and Mn in the second layer by Mn(2). 
Mn(1)-Mn(1) indicates intraplane interaction while the others interplane interactions. 
The TM-Mn(1) interaction is based on the molecular-orbital picture presented in Sec. II 
while the Mn-Mn interaction is based on the model Hartree-Fock study presented in Sec. III. 
At $d_{x^2-y^2}$ order, $d_{x^2-y^2}$ orbital is stabilized at the (interfacial) Mn layer. 
At F*, Mn-Mn interaction is weak, and the canted AF ordering may result. 
See the stabilization of the C-AF phase by the JT-type distortion $\Delta<0$ in Fig.~\ref{fig:pd} (a), 
the stabilization of the C- and A-AF phases by reducing the carrier density $N$ in Fig.~\ref{fig:pd} (b), 
and the stabilization of the A1 phase by the interfacial $d_{x^2-y^2}$ order in Fig.~\ref{fig:pd} (d). 
}
\label{tab:coupling}
\begin{center}
\begin{tabular}{ccccc} \hline \hline
$n$ & Condition & TM-Mn(1) & Mn(1)-Mn(1) & Mn(1)-Mn(2) \\
\hline 
0 & N \& JT & F & F* & F \\
  & N w/ $d_{x^2-y^2}$ order & AF & F & AF \\
1 & N & F  & F & F* \\ 
  & N w/ $d_{x^2-y^2}$ order & F &F & AF \\
  & AN & F  & F* & F* \\
  & JT w/ $N \sim 1$ & AF & F* & F \\
  & JT w/ small $N$ & F & F* & F\\
2 & N  & AF & F & F* \\
  & N w/ $d_{x^2-y^2}$ order & AF & F & AF \\
  & AN & AF & F* & F* \\
  & JT w/ $N \sim 1$ & F & F* & F \\
  & JT w/ small $N$  & AF & F* & F \\
\hline \hline
\end{tabular}
\end{center}
\end{table}

It is worth discussing the implication of the present results to the real systems. 
An example of the $(d_{3z^2-r^2})^2$ system is high-$T_c$ cuprate. 
It has been reported that the magnetic coupling between YBa$_2$Cu$_3$O$_7$ (YBCO) and 
La$_{1-x}$Ca$_x$MnO$_3$ is AF,\cite{Chakhalian06} 
and $d_{3z^2-r^2}$ and $d_{x^2-y^2}$ in the interfacial Cu have a similar amount of holes.\cite{Chakhalian07} 
This corresponds to the AN situation. 
F coupling due to the DE remains in the Mn region 
because of the finite band width of $d_{x^2-y^2}$. 
An example of the $(d_{3z^2-r^2})^1$ system is BiFeO$_3$ (BFO). 
Recently, the EB effect was reported at BFO/LSMO interfaces accompanying the ``AF'' coupling 
between BFO and LSMO.\cite{Yu09} 
We expect the N situation with $d_{x^2-y^2}$ ordering at this interface. 
Although the interfacial coupling is F, 
the AF coupling between the interfacial Mn and the second Mn layers results in the AF alignment between BFO and bulk LSMO 
as observed experimentally and is responsible for the exchange bias effect.

A question one may ask is what causes the ``AN situation'' in YBCO and the ``N situation'' in BFO? 
A qualitative explanation is as follow: 
in YBCO, the unoccupied Cu $d_{x^2-y^2}$ state is right above the Mott gap and its position is 
nearly identical to the occupied band of manganites.\cite{Yunoki07} 
Therefore, the charge transfer from Mn $e_g$ to Cu $d_{x^2-y^2}$ can easily occur. 
On the other hand, the high-spin state is realized in BFO, and 
the unoccupied $d_{x^2-y^2}$ state with opposite spin with respect to the majority electrons is located far above the gap. 
In addition, the very close chemical potentials (i.e. close $d_{3z^2-r^2}$ levels) of BFO (Ref.~\onlinecite{Paradis05}) and 
LSMO (Ref.~\onlinecite{Jong03}) maximize the B and AB splittings. 
This situation is favorable for the $d_{x^2-y^2}$ ordering in the interfacial Mn layer and 
the resulting AF coupling between the first and the second Mn layers [see Fig.~\ref{fig:pd} (d)].

Finally, an example of the $(d_{3z^2-r^2})^0$ system may be non-magnetic SrTiO$_3$, 
and the coupling with this is expected to affect the magnetic state near the interfacial Mn. 
For small doping $x$ of LSMO, 
the coupling with SrTiO$_3$ (with the smaller lattice constant of SrTiO$_3$) increases the $d_{3z^2-r^2}$ orbital occupancy
suppressing the inplane DE effect. 
For large doping, SrTiO$_3$ creates the tensile strain stabilizing $d_{x^2-y^2}$, and the out-of-plane F coupling is reduced.\cite{Ogimoto03} 
Both are expected to cause a more rapid decrease in the ordered moment with increasing temperature 
than in the bulk region,\cite{Okamoto09}
resulting in the rapid suppression of the TMR effect in LSMO/SrTiO$_3$/LSMO junctions.\cite{Bowen03,Ogimoto03}
The in-plane (out-of-plane) spin canting may also be realized in the former (latter).\cite{Ogimoto03} 
For undoped LaMnO$_3$, the out-of-plane ferromagnetic coupling may result 
because the overlap between the occupied $d_{3z^2-r^2}$ in the first Mn layer and the unoccupied $d_{x^2-z^2}$ or $d_{y^2-z^2}$ orbitals 
in the second Mn layer is increased, 
favorable for the F SE interaction between Mn layers. 

However, since $t_{2g}$ orbitals in titanates are located near (slightly above) the Fermi level of manganite,\cite{Yunoki07} 
one may need to consider $t_{2g}$ orbitals more carefully as discussed below. 

\begin{figure}[tbp] 
\includegraphics[width=0.6\columnwidth,clip]{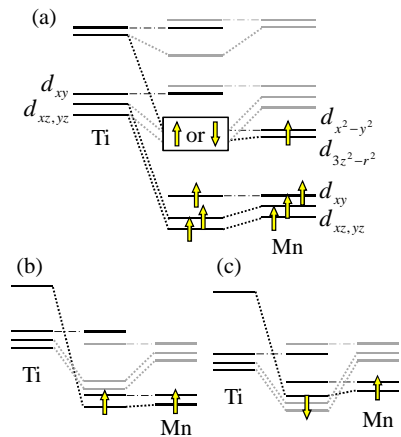} 
\caption{(Color online) Molecular orbitals formed by $3d$ orbitals on Mn and Ti, originally $d^0$. 
Here, only bonding orbitals are shown. 
(a) Full level diagram including both $e_g$ and $t_{2g}$. 
Black (light) lines indicate the level of majority (minority) spins. 
The highest occupied molecular orbitals and the magnetic alignment 
depend sensitively on the detail of the interface as shown in (b) and (c). 
(b) [(c)] The up B orbital of $d_{3z^2-r^2}$ orbital is lower (higher) in energy than 
the down B orbitals of $d_{xz,yz}$.  
In (b), induced magnetic moment in Ti is parallel to Mn, i.e., $(d_{3z^2-r^2})^0$ configuration, while  
in (c), it depends on the relative occupancy of Ti $d_{xz,yz}$ orbitals in the up and down B orbitals. 
When the occupancy of the Ti down $d_{xz,yz}$ orbitals is larger than the up $d_{xz,yz}$ orbitals, 
net moment induced in Ti site becomes antiparallel to the Mn moment. 
}
\label{fig:TiMn} 
\end{figure}

{\em Extention to $t_{2g}$ systems}. 
In $t_{2g}$ systems such as titanates, vanadates, and cromates, 
coupling between $t_{2g}$ orbitals could become as important as the coupling between $e_g$ orbitals. 
Here, we discuss how to generalize the molecular-orbital argument presented in Sec. II to $t_{2g}$ systems. 
As an example, we consider an interface between titanate with the $d^0$ configuration and manganites. 
Extending the argument to other systems is straightforward. 

Figure \ref{fig:TiMn} (a) shows the level diagram of the titanate/manganite interface including both $e_g$ and $t_{2g}$ orbitals. 
For simplicity, only bonding orbitals are presented. 
Because the $d_{3z^2-r^2}$ level in titanate is far above the occupied levels in manganite 
and the unoccupied $t_{2g}$ levels in titanate and manganite (down electrons for the latter) are close,\cite{Yunoki07} 
highest occupied molecular orbitals could be either B up $d_{3z^2-r^2}$ or $d_{x^2-y}$ orbital 
[Fig.~\ref{fig:TiMn} (b) which is equivalent to Fig.~\ref{fig:d0}]
or B down $d_{xz,yz}$ orbitals [Fig.~\ref{fig:TiMn} (c)]. 
Note that the interfacial hybridization between $d_{xy}$ orbitals on Ti and Mn is much smaller than those between $d_{xz}$ and between $d_{yz}$. 

In the case of Fig.~\ref{fig:TiMn} (b), induced moment in the titanate region is tiny but parallel to the moment in manganite region. 
On the other hand in the case of Fig.~\ref{fig:TiMn} (c), 
the induced moment in the titanate region could be either parallel or antiparallel to the manganite moment. 
This depends on the relative weight of the down electron density in the B $d_{xz,yz}$ orbitals 
with respect to the up electron density in the B $d_{xz,yz}$ orbitals. 
The situation in Fig.~\ref{fig:TiMn} (c) with antiparallel spin arrangement between titanate and manganite could happen when 
the original electron density in the manganite $e_g$ orbital is large and $d_{3z^2-r^2}$ level both in titanate and manganite is high 
due, for example, to the in-plane tensile strain. 
But in general, the difference between two configurations, Figs.~\ref{fig:TiMn} (b) and ~\ref{fig:TiMn} (c) 
with either parallel or antiparallel spin configurations, would be subtle. 
Therefore, depending on a variety of condition such as the sample preparation, any situation could be realized. 

When the number of $t_{2g}$ electrons is increased, such as doped titanates, vanadates, and cromates, 
electrons tend to enter the down B $d_{xz,yz}$ orbitals and, then, the down B $d_{xy}$, 
resulting in the antiparallel spin configuration. 
However, the antiparallel configuration becomes unstable against the parallel configuration 
when the electron number in $t_{2g}$ orbitals becomes large and 
the level separation between $t_{2g}$ and $e_g$ orbitals, i.e., $10Dq$, becomes relatively small. 

In this case, because of the strong on-site Coulomb interactions, 
the energy gain by forming B orbitals becomes small for $t_{2g}$ electrons 
and comparable to having electrons in both B and AB orbitals with the parallel spin configuration. 
The parallel configuration further lowers the energy by forming $d_{3z^2-r^2}$ B orbital and 
by the Hund coupling between the $d_{3z^2-r^2}$ B orbital and $t_{2g}$ electrons on the $t_{2g}$ system, 
i.e., the Zener's double-exchange ferromagnetism discussed in Sec.~II. 

When there are more than three $d$ electrons with relatively large $10Dq$, a low spin state is realized. 
In this case, three electrons enter B orbitals formed with $t_{2g}$ minority bands of Mn and 
remaining electrons enter AB orbitals formed with $t_{2g}$ majority band of Mn. 
Thus, the antiparallel configurations persist. 
Such a situation may be realized in, for example, an interface between manganites and SrRuO$_3$ in which Ru$^{4+}$ is in a low spin state with $t_{2g}^4$. 
This AF configuration can be turned to the F configuration when the double-exchange-type interaction becomes dominant 
due to the formation of $d_{3z^2-r^2}$ B orbital.\cite{Panagopoulos10} 

So far, we have considered the ideal lattice structure in which orbitals with different symmetry do not hybridize. 
In reality, bond angle formed by two transition metal ions and an oxygen ion in between becomes smaller than 180$^\circ$ 
allowing electrons to hop between orbitals with different symmetry. 
As a result, additional magnetic channels are generated. 
A simple argument presented in this paper can be generalized to deal with such a situation.

\acknowledgments
The author thanks P. Yu, R. Ramesh, J. Santamaria, and C. Panagopoulos 
for stimulating discussions and sharing the experimental data prior to publication, 
J. Kune{\v s} for discussion, 
and the Kavli Institute of Theoretical Physics, University of California Santa Barbara, which is supported in part by the National Science Foundation under Grant No. PHY05-51164, for hospitality. 
This work was supported by the Materials Sciences and Engineering Division, Office of Basic Energy Sciences, U.S. Department of Energy. 

\end{document}